\documentclass[runningheads]{llncs}
\usepackage{graphicx}
\usepackage{listings}
\usepackage{paralist}
\usepackage{url}
\usepackage{graphicx,multirow}
\usepackage[finalnew]{trackchanges}
\usepackage{booktabs}
\addeditor{JH}
\addeditor{EY}
\addeditor{GN}
\addeditor{OB}
\usepackage[hidelinks]{hyperref}

\begin{document}
\title{An Agent-Centric Perspective on Norm Enforcement and Sanctions}%\thanks{Supported by organization x.}}
\author{Elena Yan\inst{1}\orcidID{0009-0000-6660-9378} \and
Luis G. Nardin\inst{1}\orcidID{0000-0002-4506-2745} \and
Jomi F. H{\"u}bner\inst{2}\orcidID{0000-0001-9355-822X} \and
Olivier Boissier\inst{1}\orcidID{0000-0002-2956-0533}}
\authorrunning{E. Yan et al.}
\institute{Mines Saint-Etienne, Univ Clermont Auvergne, INP Clermont Auvergne, CNRS, UMR 6158 LIMOS, F-42023 Saint-Etienne France \email{elena.yan@emse.fr},\email{luisgustavo.nardin@emse.fr}, \email{olivier.boissier@emse.fr} \and
Department of Automation and Systems Engineering, Federal University of Santa Catarina, Florian{\'o}polis, Brazil \email{jomi.hubner@ufsc.br}}
\maketitle
\begin{abstract}
% The abstract should briefly summarize the contents of the paper in 15--250 words.
% In increasingly autonomous and distributed multi-agent systems, single centralized coordination becomes impractical and raises the necessity of governance and enforcement mechanisms from an agent-centric perspective. Sanctioning is an enforcement mechanism aimed at promoting norm compliance while preserving the agent's autonomy. To address the needs, we propose NPL(s), an extension of the Normative Programming Language NPL enriched with the representation of sanctions as first-class abstraction. Furthermore, we propose a comprehensive BDI agent architecture embedded with capabilities to manage norm enforcement and sanctions. We apply and demonstrate the interests of our approach for the control of a manufacturing process.
%\note[OB]{The abstract is fine.}
In increasingly autonomous and highly distributed multi-agent systems, centralized coordination becomes impractical and raises the need for governance and enforcement mechanisms from an agent-centric perspective. In our conceptual view, sanctioning norm enforcement is part of this agent-centric approach and they aim at promoting norm compliance while preserving agents' autonomy. The few works dealing with sanctioning norm enforcement and sanctions from the agent-centric perspective present limitations regarding the representation of sanctions and the comprehensiveness of their norm enforcement process. To address these drawbacks, we propose the NPL(s), an extension of the NPL normative programming language enriched with the representation of norms and sanctions as first-class abstractions. We also propose a BDI normative agent architecture embedding an engine for processing the NPL(s) language and a set of capabilities for approaching more comprehensively the sanctioning norm enforcement process. We apply our contributions in a case study for improving the robustness of agents' decision-making in a production automation system.
\end{abstract}
%
% ---- Sections ----
%
%==================================================================================
\section{Introduction}\label{sec:intro}
The concepts used for conceiving a multi-agent system (MAS) are often inspired by human societies, e.g., abstractions of agent, environment, interaction, and organization~\cite{boissier2020multi}.
Agents are autonomous entities that make independent decisions and act to achieve their goals~\cite{wooldridge2009mas}.
In Normative MAS (NMAS), the agents' actions are regulated by norms. A set of norms comprises an explicit and formal specification of the behaviors that agents ought to comply with for the system to achieve its overall objectives.
Agent's autonomy, however, allows agents to behave nonconforming to the prescribed norms~\cite{conte2001institutions}, which may lead the system into undesirable states.

Enforcement mechanisms are introduced to nudge agents to act conforming to the norms aiming to steer the system towards its overall objectives~\cite{hollander_current_2011}.
%
%Sanctioning is a common enforcement mechanism used to encourage agents compliance with norms.
%
% ORGANIZATION-CENTRIC PERSPECTIVE
%From an organization-centric perspective, the enforcement process is managed by an institution that monitors and controls
%
The institutional, or organization-centric, norm enforcement approach relies on a central authority to control and enforce the norms. This approach has two drawbacks. First, this approach becomes impractical in highly distributed large-scale systems. Second, it usually employs regimented enforcement mechanisms that prevent agents from violating the norms, limiting the agents' autonomy.
%
% In a social approach, normative concepts are integrated into agents to regulate their behavior. %exercise control over them. % CHECK
%
% SOCIAL / AGENT CENTRIC PERSPECTIVE
A solution to overcome these drawbacks and balance agents' autonomy and system regulation is to employ a social, or agent-centric, norm enforcement approach, in which agents supervise and regulate the behavior of others.
%, i.e., enforcement mechanisms are embedded in some or all agents.
% \note[OB]{see paper from castelfranchi on social control and discuss the difference of our approach w.r.t. trust and reputation approaches where while there are expectations on agents' behaviours and kind-of sanctions when expectations are not satisfied, norms neither sanctions are explicit.}

In our conceptual view, sanctioning is part of this agent-centric norm enforcement approach. Sanctioning is a common enforcement mechanism that uses sanctions to encourage agents to comply with norms. Sanctions are reactions to a behavior, yet not all reactions can count as sanctions~\cite{gibbs1966}.% A sanction (i) requires a referential, typically a set of norms, (ii) is applied by at least one enforcer, (iii) is associated with a prescription, (iv) specifies its enforcer's role, and (v) specifies whether it is to be perceived as a sanction by its target.  Thus, sanctions are consequences of norms violation or compliance~\cite{nardin2016sanctions}.

% Normative agents play a crucial role in adhering to and enforcing norms in MAS.
%
The NMAS community has focused on incorporating normative capabilities into agent architectures~\cite{luckEtAl2013}. In these architectures, agents should represent the norms and be able to
% ,boella2006normas,chopra2018handbook
\begin{inparaenum}[(i)]
    \item recognize new norms,
    \item recognize actions affected by the norms,
    \item reason about the norms, and
    \item decide whether or not to comply with them.
\end{inparaenum}
%~\cite{luck2013normative}
Assuming that a sanctioning enforcement mechanism is adopted, agents should also represent sanctions and be able to
\begin{inparaenum}[(i)]
    \item identify norm compliance and violation,
    \item decide for the most appropriate sanctions depending on the situation, and 
    \item apply the decided sanctions.
\end{inparaenum}

Despite the representation of norms and agents' capabilities for handling them have been well investigated in the context of NMAS, few works have proposed a comprehensive perspective on sanctioning norm enforcement from an agent-centric perspective. % Because we are interested in sanctioning enforcement mechanisms, we recognize that norms and sanctions are required as first class abstractions in any normative language.
To reduce this disparity, we advance in norm enforcement by
\begin{inparaenum}[(i)]
    \item enriching a normative programming language with the ability to represent norms and sanctions as first-class abstractions, and
    \item embedding sanctioning norm enforcement capabilities into a BDI agent architecture to enable agents to deliberate on their normative state when deciding to act.
\end{inparaenum}
%
%\note[EY]{I'm not convinced, maybe something else needs to be added in the introduction..}\note[GN]{I think the objective of the paper is an agent-centric perspective but the innovation is on the normative language. The agent architecture is an adaptation to use the extended NPL. Little is said about normative languages in the introduction. Maybe after you talk about norm enforcement architecture, you can say that for agents to reason about sanctions, this abstraction has to be available in the language and none (or few) has sanction as a first class abstraction. (Something in this lines)}%

%% CONTRIBUTION
%
%The contribution of this paper is twofold. First, we extend an existing normative language with the concept of sanction as a first class abstraction, enabling the specification of norms and sanctions for normative agents. We thus enrich its abstraction for programming norms in MAS, with those necessary for sanctions, to enable reasoning on both norms and sanctions. Second, we propose an agent-centric perspective on norm enforcement by integrating sanctioning enforcement mechanisms in a BDI agent architecture, enabling agents to also reason about, decide for, and execute sanctions.

%% PAPER STRUCTURE
The reminder of the paper is organized as follows.
In Section~\ref{sec:related}, we introduce the concepts of norms and sanctions, and discuss normative agent architectures and normative programming languages from the perspective of their support to norm enforcement.
We then extend a normative programming language by adding sanction as a first-class abstraction (Section~\ref{sec:sanction}) and extend a BDI agent architecture with sanctioning capabilities (Section~\ref{sec:architecture}).
% \note[OB]{we integrate the NPL engine that computes the sanctions to apply in the enforcement process triggered by norm violations.}
% \note[OB]{Maybe should be stressed that NPL engine has been enriched with the ability to manage the new sanction abstraction?}
%
Next, we illustrate the use of our contributions in an industrial case study (Section~\ref{sec:case-study}).
Finally, we conclude with some discussions and potential future research directions (Section~\ref{sec:conclusion}).
%
% \note[OB]{Thanks for the addition of these paragraphs. Very clear and helpful!}
% \note[OB]{Yet another question, having in mind what the COIN community could raise: to what extent isn't it an extension on a lack of the JaCaMo approach? Why did we use JaCaMo/NPL to do that? why not use other approaches?}
%==================================================================================
\section{Related Work}\label{sec:related}
%==================================================================================
% \note[EY]{

% - Norm lifecycle: adoption/ creation/enforcement/... [Elena]
%     - Sanctions [Gustavo]

% - Agent Architectures: BOID/NoA/Emilia/NBDI/ NormativeAgentSpeak [Elena]

% - Languages: NPL/2OPL [Jomi]

% - Discussion
%     Gavel [Gustavo]
% }

% \note[OB]{What this order of sub-sections? I see the logic from norm lifecycle to agent architectures w.r.t. the aim of the paper. I have difficulties to see the transition to NPL. It seems that the first two + Sanctions is a way to state the problem and motivate the work while the NPL one is more background knowledge that is used to build the proposal? If so, maybe in an other section or just cited in the technical part? }
% \note[OB]{What is Gavel? Isn't it part of Sanctions?}

%%% INTRODUCTION OF NORMS
Norm is an important concept in human societies that has been investigated in a wide range of disciplines.
%, e.g., law, sociology, psychology, economics, political sciences, social philosophy, and ethics. 
%
Norms prescribe how members of a group ought to behave in a given situation~\cite{ullmann1977}. They are expected standards of behavior socially shared and enforced by group members~\cite{hollander_current_2011}.
% conte2014minding ,singh2014norms
In MAS, norms are typically represented as obligations, permissions, or prohibitions; they are used to enable a degree of control over the autonomous agents' actions.
%
% As in this context, a norm can be conceived as a standard behavior that autonomous agents are expected to perform~\cite{hollander_current_2011,singh2014norms}. \note[GN]{Isn't this sentence redundant with the sentence above? I suggest integrating the references in the previous sentence and eliminating this one.}
%
% \remove[GN]{Norms are typically represented as obligations, permissions, or prohibitions 
% % \annote[GN]{given by a set of conditions}{What do you mean by that? Are they the activation conditions?}
% and sanctions
% Sanctions are consequences for complying with or violating them. % ~\cite{balke2013norms}
% % \change[GN]{Since agents are autonomous entities, they may decide whether or not to comply with the norms, so their actions may deviate from the expected behaviours.
% % Enforcement mechanisms can be introduced to encourage compliance by agents.}{
% Because autonomous agents' actions may deviate from the norms, enforcement mechanisms are introduced to encourage compliance.}
%}
%
%Thus, through the agents' decision-making processes, norms can be \annote[EY]{recognised, adopted, and complied with}{in lack of something better that can refer to the terms of the life cycle}, giving rise to a life cycle of the normative process.
%
%\note[OB]{The agents' decision-making processes, depend on and intervene through their decision, on the states of the norms (e.g. recognised, adopted or complied). This set of norm states defines what we call the norm's life cycle.}

The agents' decision-making processes depend and intervene through their decisions, on the state of the norms (e.g., recognized, adopted, or complied). The transitions between these norm states define what we call the norm's life cycle.

%-------------------------------------------------------------------
\subsection{Norm's life cycle}

\paragraph{\textbf{Global overview.}} There are various proposals regarding the life cycle of norms in the literature. %~\cite{finnemore_sikkink_1998,savarimuthu_review_2011,hollander_current_2011,mahmoud_review_2014}
They often share a similar structure and process but differ in specific details. Hollander and Wu~\cite{hollander_current_2011} present the norm life cycle as an overall end-to-end evolutionary process.
%
% \begin{figure}[ht]
%     \centering
%     \includegraphics[width=\textwidth]{images/norm-evolution.jpeg}
%     \caption{%\annote[OB]{The evolution process}{change the names in the figure to match the comment in the text. e.g. transmit to transmission}\note[EY]{ok!}
%     The evolution process in the norms' life cycle (based on~\cite{hollander_current_2011}).}
%     \label{fig:norm_evolution}
% \end{figure}
% %
%
%% NORM CREATION
%\noindent %\add[GN]{In Hollander and Wu's evolution process, norm's}\note[OB]{why not simply: This norm's life cycle?} 
The norm's life cycle begins with the creation of norms by system designers at design time or by autonomous normative agents at runtime (\textit{creation} phase). 
% \note[GN]{Maybe briefly explain the concept of norm emergence here.}
% \add[EY]{In the latter case, norms emerged in society from the interactions of the agents.}\note[GN]{
In the latter, norms emerge as a consequence of the agents' interactions.
%}
%
%% NORM TRANSMISSION
Once the norms exist, they are spread in the society from one agent to another through active or passive transmission (\textit{transmission} phase).
%
%% NORM ENFORCEMENT
Next, enforcement mechanisms (e.g., sanctioning) or norm compliers are used to encourage agents to adopt new norms and to comply with the adopted ones (\textit{enforcement} phase).
%
% The norm is monitored to ensure compliance through reinforcement techniques and may entail sanctions if the agent violates a norm. %\note[GN]{This seems part of the NORM ENFORCEMENT phase rather than INTERNALIZATION phase.}
%
%% NORM INTERNALIZATION
The norms adoption leads to their use in the agents' decision-making (\textit{internalization} phase).
%
%% NORM FORGETTING
As conditions change, norms may become invalid or lose their effectiveness, which cause them to be forgotten (\textit{forgotten} phase).
% \note[OB]{what do you mean by forgetting a lifecycle ? isn't it a norm?} \annote[GN]{and new ones are created through a new cycle of the evolutionary process}{This part may not be necessary because the evolutionary process is not explicit in the figure. But if you still what to talk about the evolutionary process, you should make it separate from the NORM FORGETTING, otherwise you will be entangling FORGETTING and CREATING}

Next, we describe the \emph{enforcement} phase in more detail since this paper focuses on norm enforcement.
%
%\subsection{Norm Enforcement}
%
\paragraph{\textbf{Norm Enforcement.}}\label{sec:enforcement} There are three traditional approaches to norm enforcement:
% \note[OB]{in the following replace "there is an authority" by "a single authority" ; "there are agents with" by "a subset of agents are equipped with"; }
%
\begin{inparaenum}[(i)]
  \item the \emph{institutional approach}, in which there is a single authority that enforces agents' behaviors;
  \item the \emph{social approach}, in which a subset of agents are equipped with enforcement mechanisms (e.g., police agents) used to enforce other agents' behaviors; and %\remove[GN]{ that monitor other agents};
  %\note[OB]{Is monitor the right word? they both monitor and control by interpreting the non-compliance as a violation, deciding about the sanction and applying it.}
  \item the \emph{self-enforcement approach}, in which agents enforce the norms on themselves. %\remove[GN]{ handle the consequences of the compliance or violation of the norms}.
\end{inparaenum}
%

%\note[OB]{IMO, these three items are very important. In the approach that we propose, we focus on enforcement, we don't want to use an institutional approach where the monitoring and decision to enforce is out of the agents (e.g. MOISE approach) and would like to have solutions for the two others, exploring in that paper the social one (this is why we go to the sanctions (subsection on sanction in that section and section on language for sanctions after). If so should be said somewhere.}
%
% REGIMENTED WAY AND SANCTIONS
%\note[EY]{Enforcement considers both the pre-decision and actions}
%
Cross-cutting these three approaches, the norm enforcement can be \emph{regimented} or \emph{regulated}.
% ~\cite{alechina2018norm}
In the regimented norm enforcement, agents are prevented from violating the norms, possibly limiting agents' autonomy.
%
% \change[GN]{Or through mechanisms such as sanctions in response to actions that violate norms.
% %
% In this way, agents can still perform actions that violate the norms.}{
In the regulated norm enforcement, agents can act in violation to the norms, but mechanisms, such as sanctioning, can be used to inflict consequences to their misbehavior in order to encourage their future compliance with the norms.

% SANCTIONS
In the regulated norm enforcement, sanctions can be used as a means to steer agent compliance to the norms. Sanction is a negative or a positive reaction to a violation of or a compliance with a norm.
% ~\cite{balke2013norms}
A comprehensive set of dimensions to classify sanctions and an accompanying sanctioning norm enforcement process model has been proposed in~\cite{nardin2016sanctions}. The proposed typology of sanctions enables identifying the types of sanctions that are more or less effective in reducing violation or increasing compliance. The sanctioning process model combines the institutional, the social, and the self-enforcement approaches offering opportunities for agents to reason about and decide for the most appropriate sanctions to apply in each situation.
%
% \begin{figure}[ht]
%   \centering
%   \includegraphics[width=.9\textwidth]{images/sanctioning-process-model.png}
%   \caption{Sanctioning norm enforcement process model (based on~\cite{nardin2016sanctions}).}
%   \label{fig:sanctioning-process}
% \end{figure}
%\note[OB]{use the classic JaCaMo symbolic representation of agents as used in Fig. 4? Is it possible to increase the size the figure? where is the De Facto Store?}
This model has two normative resources:
%\note[OB]{resource seems too general, what about sets or representations?}
%the \emph{De Jure} and the \emph{De Facto}.
the \emph{De Jure} that stores the representation of norms and sanctions and the \textit{De Facto} that stores the historical information about sanction decisions, executions, and outcomes. The model is enacted by five capabilities:
%\note[OB]{idem about capabilities, why not mechanisms? or components?}
%
\begin{inparaenum}[(i)]
    \item the \emph{Detector} perceives the environment and detects any action regulated by the norms in the \emph{De Jure};
    \item the \emph{Evaluator} obtains the norms and sanctions regulating the detected action from the \emph{De Jure} to determine whether the action violates or complies with
the norms. The \emph{Evaluator} also combines this information with the historical information from the \emph{De Facto} to decide the sanctions that should be applied, if any;
    \item the \emph{Executor} applies the sanctions decided by the \emph{Evaluator};
    \item the \emph{Controller} watches the regulated actions previously sanctioned to assess the efficacy of the sanctions;
    \item the \emph{Legislator} assesses the \emph{De Jure} in the light of the \emph{De Facto} information and updates the norms and sanctions in the \emph{De Jure}.
\end{inparaenum}
%

% GAVEL

%\remove[EY]{Although enforcement mechanisms are highly important, they have not been much addressed in existing architectures proposed in the literature.}\note[OB]{don't announce the conclusion of the next study here?}

Next, we review normative agent architectures (Section~\ref{sec:architectures}) and normative programming languages (Section~\ref{sec:languages}) from the perspective of their support to norm enforcement.
%\note[GN]{Not sure if we should keep ``from the perspective of their support to enforcement mechanisms''.}
%
%-------------------------------------------------------------------
\subsection{Normative Agent Architectures}
\label{sec:architectures}
%\note[OB]{24/02/13 - I am fine with the new state of this section. Thanks.}
%-------------------------------------------------------------------
%
%\note[OB]{What is the main purpose of this section w.r.t. the previous one where the life cycle of norms has been studied? If well understood, the choice of focusing on agent architecture is because we want to internalize in an agent the enforcement process and thus the analysis of existing agent architectures is focused on how norms are handled in existing agent architectures with the aim to demonstrate that there is a lack: no handling of enforcement process. Is it correct?.}
%\note[OB]{It could be helpful to see a reference to the processes in the description of the functions of the agent architecture?}
%
%
%A straightforward way to incorporate normative aspects into the agent architecture is extending the BDI (Belief-Desire-Intention)~\cite{Bratman1987-BRAIPA,rao_bdi_95} agent model with a new component for modelling norms.
%
%\change[OB]{By focusing the BDI (Belief-Desire-Intention){~\cite{Bratman1987-BRAIPA,rao_bdi_95}} agent model, a way to incorporate normative aspects in the agent architecture is extending with new components for representing and reasoning on norms.}{
Focusing on the BDI (Belief-Desire-Intention) agent model, the literature proposes normative architectures using various approaches to enforce norms.
%~\cite{Bratman1987-BRAIPA,rao_bdi_95}

% REGIMENTATION WAY: NOA, NBDI, NORMATIVE AGENTSPEAK(L)
%
Most architectures regiment norms by preventing agents from performing actions that violate them.
%
% BOID  %,broersen_boid_2002
The BOID (Beliefs-Obligations-Intentions-Desires)~\cite{broersen_boid_2001} architecture adds the concept of \emph{obligations} as a mental state that may conflict with the agent's intentions or desires. BOID agents use a predefined static priority function to resolve conflicts among these concepts. The priority ordering allows classifying agents in different categories, e.g., selfish (i.e., prioritize desires) or socially responsible (i.e., prioritize obligations).
Although BOID does not explicitly implement a norm enforcement mechanism, social responsible agents regiment norms since they always choose actions that fulfill their obligations.

Other architectures, like NoA~\cite{kollingbaum2003noa} and NBDI~\cite{dos2013developing}, regiment norms by filtering out actions or desires that violate norms. In the NoA architecture, agents use filtering functions to simply exclude actions that violate adopted norms. In the NBDI architecture, agents update their beliefs, desires, intentions and set of adopted norms independently. When selecting desires, NBDI agents only select those that are compliant with the adopted norms. Using yet a different approach, the Normative AgentSpeak(L)~\cite{meneguzzi2009normative} architecture enables agents to change their plans at runtime to conform to new norms. The approach consists in modifying the agent's plans library, i.e., create new plans when new obligation norms are adopted and suppress plans containing actions forbidden by prohibition norms. The N-Jason~\cite{lee2014njason} extends the Normative AgentSpeak(L) capabilities by scheduling intentions considering deadline and priority for runtime norm compliance.

A NoA extension~\cite{kollingbaum2006noa} and the Nu-BDI~\cite{meneguzzi2012nu} architectures relax the regimentation constraint allowing the agents to violate the norm, but diminishing the negative impact that the violation may have. In the extended NoA architecture, agents have deliberation mechanisms to annotate actions that violate norms; these annotations can be later used by the agents to decide whether or not to perform actions leading to the violation of norms. In the Nu-BDI architecture, agents annotate each step of the plans within the scope of a norm with constraints stemming from that norm, and rank the plans using a utility function based on these annotations. When selecting plans, Nu-BDI agents choose the plan that violates the norms less (i.e., highest ranked plan).

L{\'o}pez~y~L{\'o}pez et al.~\cite{lopez2006normative} implicitly incorporate regulated norm enforcement mechanisms into their agent framework. They define two levels of norms, in which the secondary norms are activated as a result of the fulfillment or violation of the primary norms. So, the secondary norms can be seen as sanctions inflicted on agents fulfilling or violating norms.

% EMIL-A
%\change[OB]{A normative component}{
%,conte2014minding
Few normative agent architectures deal explicitly with regulated norm enforcement. EMIL-A (EMergence in the Loop Architecture)~\cite{andrighetto2007emergence} is a normative BDI agent architecture developed for norm emergence, innovation, and spread. EMIL-A is endowed with modules that allow
%nardin2016emilia
% but later became a normative component. %\note[GN]{Although I used EMIL-A as a normative component, it has originally been proposed as a normative BDI agent architecture (see https://www.jasss.org/17/4/13.html).}
%
\begin{inparaenum}[(i)]
    \item norm recognition (i.e., recognition and representation of norms as mental states),
    \item norm adoption (i.e., detection and dynamic updating of the norms salience that corresponds to the perceived prominence of a norm within a relevant reference group),
    \item norm compliance (i.e., decision to comply with the adopted norms), and
    \item norm enforcement (i.e., detection of norm compliance and violation, and the application of sanctions).
\end{inparaenum}
The n-BDI architecture~\cite{criado2014reasoning} also incorporates a norm enforcement mechanism that always rewards or punishes the agents as a consequence to, respectively, the compliance with or violation of a norm. %The normative mechanism uses norm salience for the norm acquisition and for determining norm relevance.
AORTA~\cite{jensen2015aorta} proposes a normative agent module that provides normative reasoning capabilities to the agents. These agents are able to reason about norms, automatically detect violations as soon as they become aware of the violation state, and produce a new belief about the violation that may trigger another norm or plan.

%%% NON-BDI ARCHITECUTRES: MDP
In addition to those BDI-based normative agent architectures, Fagundes et al.~\cite{fagundes2010mdp} extend the Markov Decision Process (MDP) framework for creating self-interested agents able to reason about norms. The framework uses an enforcement mechanism based on the imperfect observation of the system state and agents, and inflicts a cost on agents if a violation is detected~\cite{fagundes2014enforcement}. Agents make decisions based on the expected utility of conforming with or violating a norm, which takes into account the cost of violating the norm and the probability that the norm violation will be detected.
%
%-------------------------------------------------------------------
\subsection{Normative Programming Languages}\label{sec:languages}
%-------------------------------------------------------------------
Normative programming languages provide abstractions that enable agents to deliberate about their actions bounded by the normative regulations. These languages complement the normative agent architectures as they provide an explicit representation of and enable the handling of normative concepts (e.g., obligation, permission, prohibition, sanction). There has been extensive work on normative programming languages with an organization-centric perspective~\cite{dastaniEtAl2009,hubner:09c}; however, there has been limited interest in languages with an agent-centric perspective.
% riemsdijkEtAl2009, dybalovaEtAl2014, padgetEtAl2016

One of the first normative programming languages with an agent-centric perspective is the NoA language~\cite{kollingbaum2002noa}. The language, whose semantics are implemented by the NoA architecture, contains constructs for the specification of beliefs, goals, plans and norms. Norm specifications can regulate states (i.e., the achievement of a particular state of the world) or actions (i.e., the performance of explicit actions). In this language, normative statements are expressed as obligations, permissions, prohibitions, and sanctions. Although sanction exists as an independent construct, it is a syntactic sugar that sets an obligation for an agent to pursue certain activities that represent such sanctions.

The Normative Programming Language (NPL)~\cite{hubner:11a}, albeit used mainly within the organization-centric perspective~\cite{hubner:09c}, is a general language dedicated to the development of normative programs based on two primitive constructs: obligation and regimentation. Other constructs like prohibition, permission, and sanction are represented using these primitive constructs. Prohibitions are accomplished by regimentation or by delegating to someone an obligation to deal with the situation. Permissions are defined by omission, i.e., NPL adopts the open-world assumption. Likewise the NoA language, the NPL also represents sanctions as obligations.

The N-2APL~\cite{alechinaEtAl2012} is an extension to the agent programming language 2APL~\cite{dastaniEtAl2009}. The N-2APL provides support for representing beliefs, goals, plans, norms (i.e., obligations or prohibitions), sanctions, deadlines, and durations. A sanction is a consequence to the violation of an obligation or prohibition, i.e., specify the updates to be applied on the environment due to the violation of a norm. In N-2APL, sanction is part of the norm specification, thus tightly associated to it.
%
%-------------------------------------------------------------------
\subsection{Remarks}\label{sec:remarks}
%-------------------------------------------------------------------
In our conceptual view, sanctions are part of social mechanisms (i.e., agent-centric perspective) for norm enforcement as they are consequences of norms violation or compliance. Although sanctioning mechanisms are built on top of norms and their state, sanctions are not simply triggered depending on norms state, but also depending on the context in which the compliance or violation take place. Tackling these requirements call for a normative programming language that disentangles sanctions representation from, yet letting it associated to, norms, and a normative agent architecture that triggers sanctions from a normative state contingent upon the agent's contextual situation in order to fulfill the agent-centric perspective.

In the next sections, we propose an extension to the NPL~\cite{hubner:11a} normative language and to the JaCaMo agent~\cite{bordini:07} to tackle these requirements. We chose to extend NPL because, to the best of our knowledge, it is the only general purpose normative programming language available, and the JaCaMo agent because NPL has already been integrated into it in an organization-centric perspective~\cite{hubner:09c}.

% a general view of the proposal
%A sanction mechanism is thus built on top of norms and their state. For example, the sanction ``fine Alice \$100'' might be the consequence of violating the norm ``every agent is forbidden to cross red traffic lights.'' Hence a sanction is created, it can be implemented in several and domain dependent ways. By ``implement'' we mean, for instance, how the fine is concretely realised in an application and by who. Our proposal in this paper focuses on the creation of sanctions from a normative state and does not address their implementation. 
%\note[OB]{are the word "created" and "implemented" the best ones? is it the way people talk about enactment and execution of a sanction in the literature?}

%\note[OB]{This first sentence should be the conclusion of the previous section. Maybe should be extended to clearly state what is our conceptual view, including agent-based enforcement (including monitoring which is externalized and handled in a semi centralized way in the current JaCaMo) with an explicit decision about the sanctions to apply).}
%==================================================================================
\section{NPL(s): Extending NPL with Sanctions}\label{sec:sanction}
%==================================================================================
% NPL and Integration of Norm Sanctions
% formal definition and specification
%
% BNF
\newenvironment{bnfgrammar}{\begin{tabbing}\bnfnt{formula} \= \texttt{\small ::=} \= ~~~ \= \kill}{\end{tabbing}}
\newcommand{\bnfline}[2]{\bnfnt{#1} \>\texttt{\small ::=} \>#2}
\newcommand{\bnfnt}[1]{\textrm{\textit{#1}}} %$\langle$#1$\rangle$}
\newcommand{\bnfl}[1]{``\texttt{\textbf{#1}}''}
\newcommand{\bnfor}{$\mid$~}
%
% a general view of the proposal
%A sanction mechanism is thus built on top of norms and their state. For example, the sanction ``fine Alice \$100'' might be the consequence of violating the norm ``every agent is forbidden to cross red traffic lights.'' Hence a sanction is created, it can be implemented in several and domain dependent ways. By ``implement'' we mean, for instance, how the fine is concretely realised in an application and by who. Our proposal in this paper focuses on the creation of sanctions from a normative state and does not address their implementation. 
%\note[OB]{are the word "created" and "implemented" the best ones? is it the way people talk about enactment and execution of a sanction in the literature?}
%
% justify a bit the approach of a language
% \note[OB]{here begins the contribution. This should be clearly said. Come back to the objective in relation to the current limitations of the literature.}
%
Here we extend the NPL~\cite{hubner:11a} language to introduce the concept of sanctions as a first-class abstraction. Our approach consists of having an enriched and separated description in formal language of sanctions and the conditions that allow them to be triggered. This language is interpreted at run-time based on the current state of the normative system. In addition, the NPL interpreter should be extended to compute new sanctions. The new language, named NPL(s), is thus presented in the sequel by its syntax and (informal) semantics.

% \change[OB]{A normative program \bnfnt{np} in NPL is composed of: ($i$) a set of facts and inference rules (based on a syntax similar to the one used in
% Jason~\cite{bordini:07}); and ($ii$) a set of norms.  NPL(s) extends it with ($iii$) a set of sanction rules.}{
A normative program \bnfnt{np} in NPL(s) is composed of a set of: ($i$) facts and inference rules (based on a syntax similar to the one used in Jason~\cite{bordini:07}), ($ii$) norms, ($iii$) sanction rules. While the two first components come from NPL, the last one is the extension introduced by NPL(s).
We briefly introduce NPL norms in this paper, but the formal semantics and more details are available at~\cite{hubner:11a}. An NPL norm has the general form:
\begin{center}
  \texttt{norm $\mathit{id}$ : $\varphi$ -> $\psi$ .}
\end{center}
where $\mathit{id}$ is a unique \emph{identifier} of the norm; $\varphi$ is a formula that determines the \emph{activation condition} for the norm; and $\psi$ is the \emph{consequence} of the activation of the norm. Among the possible consequences, here we focus on obligations (other possible consequences are permissions, prohibitions and failures).

\begin{itemize}
  \item[] $\psi = \mathtt{obligation}(a,m,g,d)$: represents the case where an obligation for some agent $a$ is created.  Argument   $m$ is the maintenance condition for the obligation; $g$ is the formula that represents the obligation itself (a state of the world that the agent must try to bring about, i.e., a goal it has to achieve); and $d$ is the deadline condition to fulfill the obligation.
\end{itemize}
\noindent
The informal semantics defines that whenever $\varphi$ is true, an obligation for $a$ is \add[JH]{created and its initial state is} \texttt{active}. If $m$ becomes false, the obligation \add[JH]{state changes to} \texttt{inactive}. If $g$ becomes true \add[EY]{before $d$ is true}, the obligation \add[JH]{state changes to} \texttt{fulfilled}. If $d$ becomes true \add[EY]{before $g$ is true}, the obligation \add[JH]{state changes to} \texttt{unfulfilled}. %\annote[OB]{Inactive, fulfilled, and inactive are the final states of an obligation}{
The final state of an obligation may thus be \texttt{unfulfilled}, \texttt{fulfilled}, or \texttt{inactive}.

\begin{table}[!tb]
\centering
\rule{\linewidth}{.5pt}
\begin{bnfgrammar}
\bnfline{np}{\bnfl{np} \bnfnt{atom} \bnfl{\{}  ( \bnfnt{rule} \bnfor \bnfnt{norm} \bnfor \bnfnt{srule} )* \bnfl{\}}} \\
  \bnfline{rule}{ \bnfnt{atom} [ \bnfl{:-} \bnfnt{formula} ] \bnfl{.}}\\
  \bnfline{norm}{\bnfl{norm} \bnfnt{id} \bnfl{:} \bnfnt{formula} \bnfl{->}  (
    \bnfnt{fail} \bnfor \bnfnt{obl} \bnfor \bnfnt{per} \bnfor \bnfnt{pro} )  (\bnfnt{tsr})* \bnfl{.}}\\[.2cm]
 \bnfline{fail}{\bnfl{fail(}  \bnfnt{atom} \bnfl{)}} \\
  \bnfline{obl}{\bnfl{obligation} \bnfnt{dargs} }\\
  \bnfline{per}{\bnfl{permission} \bnfnt{dargs} }\\
  \bnfline{pro}{\bnfl{prohibition} \bnfnt{dargs} }\\
  \bnfline{dargs}{\bnfl{(} (\bnfnt{var} \bnfor \bnfnt{id}) \bnfl{,} \bnfnt{formula} \bnfl{,}  \bnfnt{formula} \bnfl{,} (\bnfnt{time} \bnfor \bnfnt{formula}) \bnfl{)}}\\[.2cm]
  
  \bnfline{tsr}{\bnfl{if} (\bnfl{fulfilled} \bnfor \bnfl{unfulfilled} \bnfor \bnfl{inactive}) \bnfl{:} ( \bnfnt{atom} )*}\\[.2cm]

   \bnfline{srule}{\bnfl{sanction-rule} \bnfnt{atom} \bnfl{:} \bnfnt{formula} \bnfl{->} \bnfl{sanction(} (\bnfnt{var} \bnfor \bnfnt{id}) \bnfl{,} \bnfnt{atom} \bnfl{)}}\\[.2cm]

  \bnfline{formula}{\bnfnt{atom} \bnfor \bnfl{not} \bnfnt{formula} \bnfor  \bnfnt{atom} ( \bnfl{\&} \bnfor \bnfl{|}) \bnfnt{formula}}\\
  \bnfline{time}{ \bnfl{`} \bnfnt{number}   (  \bnfl{second}  \bnfor \bnfl{minute} \bnfor ...) \bnfl{`}}
\end{bnfgrammar}
\rule{\linewidth}{.5pt}
\caption{EBNF of the Normative Programming Language for Sanctions ---
  NPL(s). Non-terminals \bnfnt{atom}, \bnfnt{id}, \bnfnt{var}, and
  \bnfnt{number} correspond, respectively, to predicates, identifiers,
  variables, and numbers as used in Prolog.\label{fig:syntaxnpl}}
\end{table}

In NPL(s), a norm can be followed by sanction rules to be evaluated according to the final state of an obligation (see \bnfnt{tsr} in \change[EY]{Fig.}{Table}~\ref{fig:syntaxnpl}):

\begin{flushleft}
  \texttt{norm $\mathit{id}$ : $\varphi$ -> $\psi$}\\
  \mbox{}\hspace{1cm}\texttt{if fulfilled: $sr_1(args), sr_2(args), ... , sr_n$ .}
\end{flushleft}

\noindent
A sanction rule has the following general form (see~\bnfnt{srule} in \change[GN]{Fig.}{Table}~\ref{fig:syntaxnpl}):
\begin{flushleft}
    \texttt{sanction-rule $\mathit{sr_i}$($\mathit{args}$) : \change[EY]{$\varphi$}{$\rho$} -> sanction($a$, $s$) .}
\end{flushleft}

\noindent where $\mathit{sr_i}$ is a unique \emph{identifier} of the sanction rule; $\mathit{args}$ are terms passed as parameters (optional); \change[EY]{$\varphi$}{$\rho$} is a formula that determines the \emph{activation condition} for the sanction (optional, assumed true if not specified); $a$ is the target agent of the sanction; and $s$ is the sanction content. Sanction rules can be read as ``In the \change[JH]{context of $sr_i(args)$}{case that a norm has triggered the sanction rule $sr_i$ because it is (un)fulfilled or inactive}, if \change[EY]{$\varphi$}{$\rho$} is true, then the sanction $s$ is created for agent $a$.''
%%\change[OB]{Sanction rules can be read as ``if $sr_i$ is triggered by a final state of some obligation and the condition $\varphi$ is true, then the sanction $s$ is created for agent $a$.''}{Sanction rules can be read as ``In the context of $sr_i$, if $\varphi$ is true, then the sanction $s$ is created for agent $a$.''}
%\note[OB]{I just changed the sentence to be consistent with the reading of a norm expression as done above. norm id is a way to type a norm and to give an id. For sanctions it is the same sanction-rule is a type, sr is the id of the sanction rule.}
%\note[OB]{Where are the $args$ in the formula defining the sanction?}
A simple example to illustrate the language is given below\change[EY]{; we use source code comments to explain the program.}{. Two norms are defined, each associated with a sanction-rule. Norm \texttt{n1} specifies that \texttt{alice} is obligated to achieve \texttt{b(0)} within 3 seconds. Unfulfillment to do so activates sanction-rule \texttt{sr1}, resulting in the imposition of a \texttt{fine} to \texttt{alice}. If the sanction is created and the condition \texttt{extra(C)} is met, norm \texttt{n2} is activated. This delegates \texttt{bob} to apply the fine to \texttt{alice} in 2 seconds. In the event of non-compliance, a sanction is issued to remove \texttt{bob} from the system.}

% %
\begin{quote} {\footnotesize
\begin{verbatim}
// * NORMS *
// alice has 3 seconds to achieve b(0), or else evaluate sr1
norm n1: vl(X) & X > 5
   -> obligation(alice,true, b(0), `3 seconds`)
      if unfulfilled: sr1(alice,X) .

// bob is obliged to apply fines in 2 seconds
norm n2: sanction(A,fine(X)) & extra(C)
   -> obligation(bob,true, apply_fine(A,X*C), `2 seconds`)
      if unfulfilled: sr2 .

// * SANCTION RULES *
// if A is not in an emergency, create fine sanction for it
sanction-rule sr1(A,V) : not emergency(A) -> sanction(A,fine(V)) .

sanction-rule sr2 -> sanction(bob,remove_from_systems) .
\end{verbatim}
}
\end{quote}

%\note[OB]{last sanction-rule seems to miss a \texttt{: true} in the condition? I suppose it is not required when is true, but I included just in case.}
%\note[EY]{the condition in the sanctions is optional, it should be specified previously in the definition} DONE

\noindent
Based on this NPL(s) example, we can have the following example story line:
\begin{enumerate}
    \item A fact \texttt{v(20)} is produced and norm \texttt{n1} is triggered.
    \item \texttt{alice} is thus obliged to achieve \texttt{b(0)} in 3 seconds.
    \item After 3 seconds, \texttt{alice} does not fulfill the obligation, triggering the sanction rule \texttt{sr1(alice,20)}.
    \item Since \texttt{alice} has no fact like \texttt{emergency(alice)}, the sanction rule \texttt{sr1} produces the \texttt{sanction(alice,fine(20))} fact. 
    \item This new sanction fact together with the fact \texttt{extra(10)} triggers the norm \texttt{n2} that obliges agent \texttt{bob} to \texttt{apply\_fine(alice,200)} in 2 seconds. 
    \item Supposing that \texttt{bob} fulfills its obligation, the story ends here.
    \item However, if \texttt{bob} does not fulfill the obligation, the sanction rule \texttt{sr2} is triggered and the \texttt{sanction(bob,remove\_from\_systems)} fact is produced.
    %\add[EY]{In this case, it is assumed that  %\change[JH]{there is an artifact from the environment responsible for handling }{the environment is capable to handle} the sanction and apply it.}
    \add[EY]{Here, we assume the environment is capable of handling and applying the sanction.}
\end{enumerate}

\noindent
From this story, we can notice that the NPL(s) interpreter is limited to compute sanction \add[JH]{facts}. The impact of these sanctions on the running system depends on something else that reads the sanction fact and implements it.
\add[EY]{This could be integrated in an \emph{agent-}, \emph{environment-}, or \emph{organization-}centric perspective.}
% agent-centric
For instance, it could be an agent that reads the sanction fact and \emph{decides} to run a procedure to implement it. 
% environment
However, it could be the case that the environment is capable of reading sanction facts and implementing them, as it is the case of automatic fines created when we cross red traffic lights. 
% npl interpreter
Finally, the NPL(s) interpreter itself can read sanction facts, as in the example of the activation of norm \texttt{n2}.
%
%\note[OB]{using example of a speed limit could be more explicit with a radar with the automated production of official fine}. 
%\note{for extended paper (after workshop) we can include formal semantics and interpreter algorithm.}
%\note{I did not include count-as or enact-as since we plan to use an inside agent approach}
%
\add[EY]{In this paper we focus on the agent-centric perspective.}
%==================================================================================
\section{Normative Agent Architecture}\label{sec:architecture}
\begin{figure}[ht]
      \centering
      \includegraphics[width=0.7\textwidth]{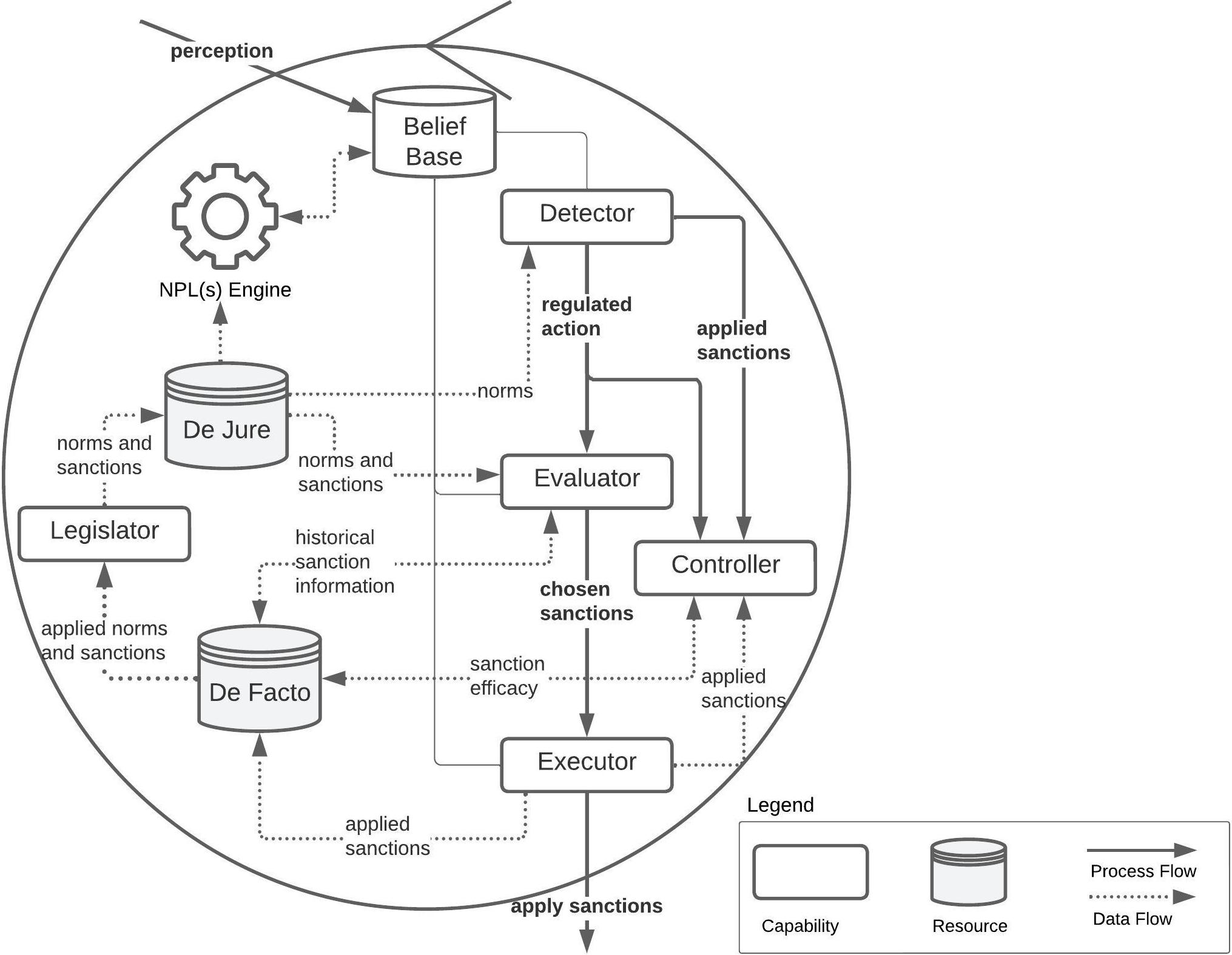}
      \caption{Normative BDI agent architecture extended to handle normative states.\protect\footnotemark}
      \label{fig:agent-architecture}
\end{figure}

%==================================================================================
% \note[EY]{
% - differences: centralised perspective/decentralised perspective, 
% - agent reasoning about norms/capabilities, 
% - normative engine / de jure
% }
%
% REMARK: modular architecture
% REMARK: different agents could have different reactions to the same sanction (shows the flexibility)
Here we describe an extension of the JaCaMo BDI agent architecture to handle normative states as illustrated in Fig.~\ref{fig:agent-architecture}\footnote{The source code of the normative agent architecture is available at~\url{https://github.com/moise-lang/npl/blob/master/src/main/java/npl/NormativeAg.java}}.
%
%\note[EY]{Fig.2 to be checked, the shared diagram link: https://t.ly/Ti8S5}
\footnotetext{Note that this architecture depicts only norm enforcement capabilities. Others were omitted due to our interest in norm enforcement in this paper.}

This normative architecture integrates an NPL(s) Engine enabling agents to interpret and compute the current state of norms and sanctions specified in the NPL(s) language (Section~\ref{sec:languages}). Additionally, the normative architecture draws inspiration from the sanctioning norm enforcement process model proposed in~\cite{nardin2016sanctions} (Section~\ref{sec:enforcement}) to define the specialized resources and enforcement capabilities needed to implement a normative agent. Although these capabilities are present in the architecture, they are implemented as plans that are triggered by normative facts. This characteristic gives modularity and flexibility to agents implementing this architecture as
\begin{inparaenum}[(i)]
  \item not all capabilities have to be implemented in each agent and
  \item different agents can implement these capabilities differently, i.e., agents can deal differently with norms and sanctions.
\end{inparaenum}
An exception is the Detector capability, which, in practice, is performed by the NPL(s) Engine.

The De Jure is a normative program, i.e., norms, sanction rules, and sanctions, specified in NPL(s) language. The specification is read once and used by the NPL Engine to together with the belief base information of the agent determine the current state of norms and sanctions, as well as add normative facts in the belief base. Once added into the belief base, these normative facts may trigger specific capabilities (i.e., plans). For instance, once a norm is activated, an obligation is added to the belief base causing the execution of a plan; or once norm is unfulfilled, a sanction rule is triggered producing sanction facts in the agent's belief base. Contrary to the De Jure, the De Facto is part of the belief base of the agent that contains particular types of beliefs, i.e., beliefs related to the decision and application of sanctions and their efficacy.
%\note[EY]{connection between the belief base and the detector capability?}
%\note[OB]{De Facto should be part of the belief base: this is a particular type of beliefs}
%
% Detector is played by NPL engine
%
% TODO: integrate npl(s), npl engine, belief
%...

%==================================================================================
\section{Case Study: Laboratory Plant MyJoghurt}\label{sec:case-study}
%==================================================================================
% \note[EY]{Reference paper: ``Increasing Robustness of Agents’ Decision-Making in Production Automation using Sanctioning"
%
% - problem
%
% - architecture
%
% - prototype
% }
%
Here we present a case study illustrating the practical potential of the extended normative programming language NPL(s) integrated into the proposed normative agent architecture. %\change[OB]{in improving the robustness of agents' decision-making in a production automation system}{. 
The use case is focused on improving the robustness of agents' decision-making in a production automation system.
%
%\chnage[OB]{We took as a reference the industrial use case of the laboratory plant \textit{myJoghurt} introduced in~\cite{land2023automation}.
%The laboratory plant \textit{myJoghurt} }{
We took as reference the industrial use case of the laboratory plant \textit{myJoghurt} introduced in~\cite{land2023automation}.

\begin{figure}[htb]
      \centering
      \includegraphics[width=0.6\textwidth]{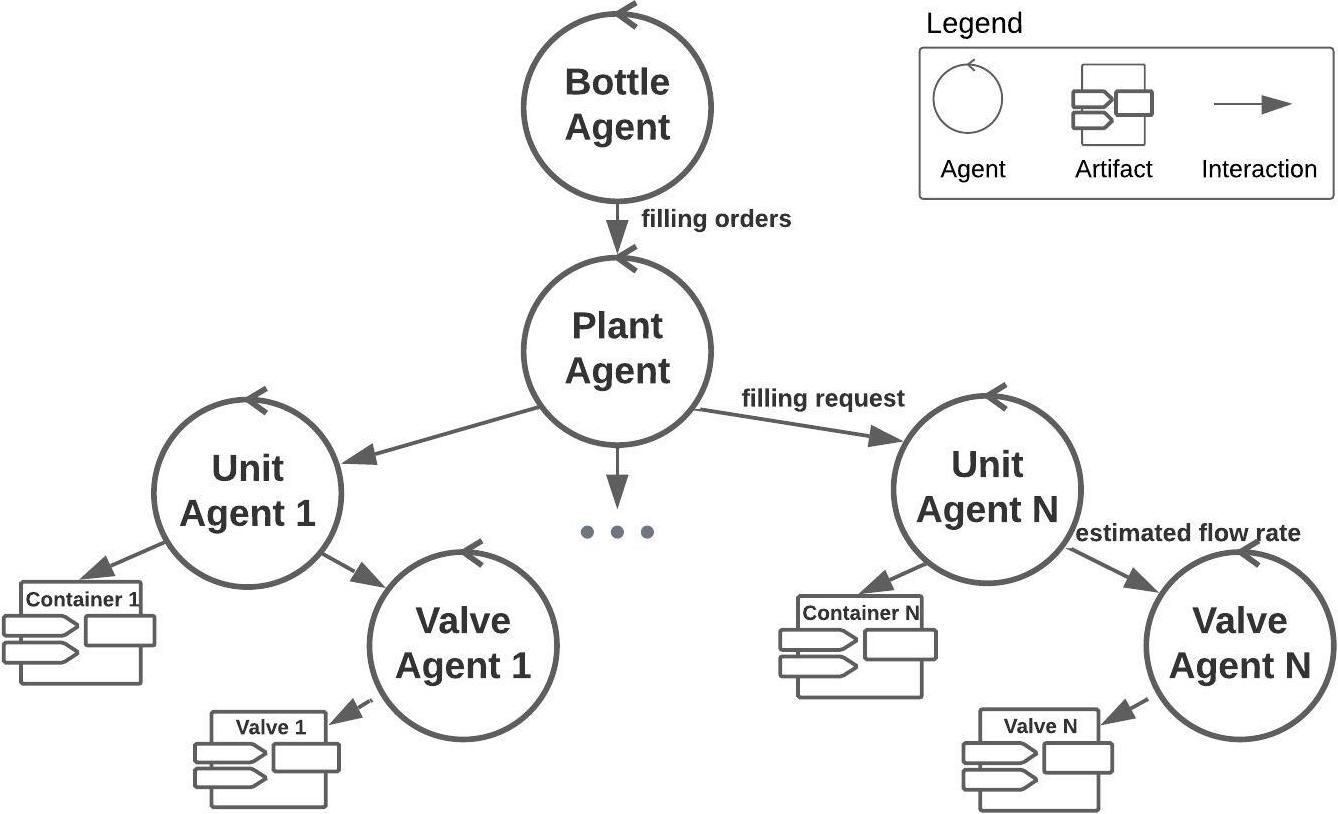}
      \caption{\emph{myJoghurt} multi-agent architecture (based on \cite{land2023automation}).}
      \label{fig:architecture-isa88}
\end{figure}

The \textit{myJoghurt} is composed of filling stations responsible for filling recipe-specific liquids (e.g., yogurt, milk) into bottles that are transported by the logistics system. Orders are received and the filling tasks are distributed among the process plant.
The orders and filling process are controlled by a multi-agent system (illustrated in Fig.~\ref{fig:architecture-isa88}), whose agents control the process plant into filling the correct amount of liquid in the bottles according to the order received.
%\note[OB]{Add a legend to explain symbols of agent and artifact}

\noindent
The \textit{myJoghurt} control system is composed of five types of components: bottle, plant, unit, container, and valve.
%
%
%\footnote{The source code of the case study is available at~\url{https://github.com/yan-elena/myjoghurt-case-study}}
%The technology that we used for agentifying this system is JaCaMo~\cite{boissier2020multi} with the integration of NPL(s) for the normative and sanction parts.
%The resulting multi-agent architecture of the process plant is illustrated in Fig.~\ref{fig:architecture-isa88}.
%
They are controlled by four types of agents that interact with each other and two types of artifacts. The \textit{bottle agent} receives and sends filling orders from customers with the type and the amount of liquid to the \emph{plant agent}.
The \textit{plant agent} manages the process at the cell level and negotiates with the \textit{unit agent} when orders are received.
The \textit{plant agent} then selects and assigns the order to \textit{unit agents} based on their availability.
The \textit{unit agent}, which controls the filling station, interacts with the filling station's container and valve artifacts for the filling process.
%
% \annote[OB]{The \textit{container agent} can contain different liquids (such as yoghurt, milk, or others)}{not so clear what is an agent that contains? manages the level of liquids in the container?}, and the filling process is handled by a \textit{valve agent} that opens for a specific time to fill the liquid in the bottle.
% \change[EY]{The \textit{container} can contain different liquids (such as yogurt, milk, or others)}{
The \textit{container}, represented as an artifact, manages the level of liquids in the container.
The \textit{valve agent} estimates the flow rate considering the type of liquid (e.g., a liquid with a higher viscosity leads to a faster flow) and whether the valve is clogged (e.g., residues of liquid could clog the valve over time).
Based on the estimated value, the \textit{valve agent} controls the opening window of the valve through the \textit{valve} artifact.

%%% ENFORCEMENT MECHANISMS AND SANCTIONS
Enforcement mechanisms and sanctions are used to steer the decisions of agents to improve the accuracy of the flow rate estimation in various types of liquid and in the presence of non-deterministic situations.
% NORM
The norms defined in this use case refer to ensuring the correct fulfillment of order requests of each bottle (i.e., bottles filled with the correct liquid and volume within a specified tolerance range).
% FACTORS
The system considers the deviation and learning factors to determine whether and which sanctions to apply.
The deviation factor comprises polarity, determined by whether an action is complied with (positive polarity) or violated (negative polarity), and magnitude that measures the degree of deviation from the expected fill level.
The learning factor considers the historical behavior of the \textit{valve agents}, represented by
%\note[OB]{again a factor. Is it in the learning factor or a new factor of the evaluator model?}
the image of the agent (i.e., assessment based on direct experience with the agent~\cite{sabater2006repage}), the frequency of violations, and the efficacy of sanctions applied. %\note[OB]{aren't all these information part of the De Facto?}
%
% SANCTIONS
% If the measured fill level is in an approved range, the polarity of the deviation factor is set to positive, and the unit agent may receive a positive sanction, which increases its image.
% If instead, the measured fill level is outside the approved range, the image of that liquid will decrease. 
% Based on the updated deviation and learning factors, four kinds of sanctions may be chosen.
% \remove[EY]{They decide the sanction(s) to apply according to the information stored in De Jure (i.e., all norm and applicable sanctions) and De Facto (i.e., historical sanction information and applied sanctions).\note[OB]{From the schema not clear if this is in an agent or out.}}
%
% The Evaluator capability of unit agents may decide to apply sanctions S1 and/or S2 to their associate valve agent:
%\note[OB]{To what extend are they sanctions? seems to be exception handling actions rather than sanctions?}

\add[EY]{Table{~\ref{tab:sanctions}} illustrates the sanctions defined for the case study. The activation conditions of sanctions are evaluated based on the compliance or violation of the norm, and sanctions are applied accordingly. 
All sanctions have a negative polarity{~\cite{nardin2016sanctions}}, intended to punish the \textit{valve} or the \textit{unit agent} for their misbehavior. S1 aims to guide the \textit{valve agent} by adjusting its estimated flow rate, while all other sanctions aim to reduce (e.g., S3) or prevent future misbehavior temporarily (e.g., S2 and S5) or permanently (e.g., S4).
}

\note[EY]{another word better than \textit{sanctioner}, which makes the table more compact?}

\begin{table}[htb]
\centering
%\begin{tabular}{|p{0.15\textwidth}|p{0.09\textwidth}|p{0.36\textwidth}|p{0.36\textwidth}|}
\renewcommand{\arraystretch}{0.95}
\begin{tabular}{lllp{0.35\textwidth}p{0.35\textwidth}}
\toprule
\textbf{Id} & \textbf{Sanctioner} & \textbf{Target} & \textbf{Sanction} & \textbf{Pre-condition} \\ \midrule
S1 & unit & valve & Adjust the estimated flow rate & The image is below a threshold \\
\hline
S2 & unit & valve & Activate the self-cleaning procedure & The violation occurs three consecutive times \\
\hline
S3 & plant & unit & Adjust the unit agent's image & The image is below a threshold \\
\hline
S4 & plant & unit & Disregard the unit agent as an option for subsequent filling orders & The violation occurs five consecutive times \\
\hline
S5 & plant & unit & Require manual intervention to repair the filling station & The violation occurs five consecutive times \\
\bottomrule
\end{tabular}
\caption{Sanctions defined for the case study. The sanctioner agent applies the sanction to the target agent if the activation pre-condition holds.}
\label{tab:sanctions}
\end{table}
\note[EY]{changed the list of sanctions in a table}
%
%\begin{enumerate}
\remove[EY]{
    (S1) adjust the estimated flow rate if the image is below a threshold;
    (S2) activate the self-cleaning procedure if the problem persists for two or three consecutive times.
}
%\end{enumerate}
%
% According to the classification of sanctions~\cite{nardin2016sanctions}, S1 and S2 are considered directed, as they are addressed to the valve (i.e., other-directed locus and direct mode).
% The purpose of these sanctions is to guide the valve to improve flow estimation (i.e., guidance performance purpose).
%
\remove[EY]{
The \textit{plant agent} may apply the following sanctions to the \textit{unit agent}:
%
%\begin{enumerate}
    (S3) adjust the \textit{unit agent}'s image, if the image is below a threshold, the likelihood of selecting the \textit{unit agent} is reduced;
    (S4) disregard the \textit{unit agent} as an option for subsequent filling orders if a violation persists for five consecutive times;
    (S5) generate an alarm and require manual intervention to repair the filling station if a violation persists for five consecutive times. %indirected
}
%\end{enumerate}
%
% Sanctions S3 and S4 are applied to the plant agent itself (i.e., self-directed locus) to update the unit agent's image due to the latter's misbehavior (i.e., indirect mode).
%
% Similarly, the sanction S5 is a sanction directed to others and operates indirectly (i.e., other-directed locus and indirect mode).
%
% These decisions influence the future order assignments to the unit agent (i.e., incapacitation performance purpose) and it is unaware of these decisions (i.e., unnoticeable discernability).
%
\noindent
We used JaCaMo~\cite{boissier2020multi} with the extended NPL(s) normative language to implement the case study\footnote{The source code available at~\url{https://github.com/yan-elena/myjoghurt-case-study}}.
%The resulting multi-agent architecture of the process plant is illustrated in Fig.~\ref{fig:architecture-isa88}.
%
%---------------------------------------------------------
\subsection{Representation of Norms and Sanctions}
%---------------------------------------------------------
%
%\note[EY]{If we exceed the number of pages, we can summarize this section}
%In this section, we illustrate a part of the implementation\note[OB]{what is the coverage of the case study or simplifications? seems that agents don't implement all elements of sanctioning mechanism?} of the case study.
%
Here, we present how norms and sanctions are represented in the De Jure repository using the NPL(s) normative language (see Section~\ref{sec:sanction}).
%
% Each De Jure repository with norms and sanction rules is implemented in the NPL(s) language.
%
The listing below illustrates the %\change[EY]{implementation of the De Jure of the unit agent.}{%\note[OB]{Is it the implementation or simply the 
\textit{Unit De Jure} repository with the representation of the \textit{unit agent}'s norms, sanction rules, and sanctions\footnote{The \textit{plant agent}'s norms, sanction rules, and sanctions are available at~\url{https://github.com/yan-elena/myjoghurt-case-study/blob/main/src/npl/plant_de_jure.npl}}.

\begin{lstlisting}[
    language={},
    label={lst:unit_norms}
]
norm n1: fill_bottle(LQ, X, MN, MX) & .my_name(U)
    -> obligation(U, n1, fill(LQ,X,MN,MX), level(X,L) & (L<MN | L>MX)).

norm n2: level(V,X,L) & .my_name(U)
    -> obligation(U, n2, update_factors(V,X,L), deviation_factor(V,X,"negative",_) & learning_factor(V,X,I,_,_,_) & threshold(T,_) & I<T)
    if unfulfilled: s1(V,X), s2(V,X).

sanction-rule s1(V,X): learning_factor(V,X,_,_,_,C) & threshold(_,T) & C<T
    ->  sanction(V, adjust_flow_rate(X)).

sanction-rule s2(V,X): learning_factor(V,X,_,_,_,C) & threshold(_,T) & C>=T
    -> sanction(V, self_cleaning(X)).
\end{lstlisting}

%\note[OB]{here we have the functioning of the NPL(s) engine.} 
The working principle of the multi agent-centric sanctioning mechanism including the NPL(s) engine operates as follows. The NPL(s) activates the norm \texttt{n1} whenever the \textit{unit agent} receives a filling request \texttt{fill\_bottle(LQ, X, MN, MX)} from the \textit{plant agent}.
%\note[OB]{I would reformulate as: The working principle of the multi agent-centric sanctioning mechanism including the NPL(s) engine is as follows.}
%
%when the \textit{unit agent} receives a filling request \texttt{fill\_bottle(LQ, X, MN, MX)} from the \textit{plant agent}, the norm \texttt{n1} is activated.
%
This norm produces an obligation \texttt{fill(LQ, X, MN, MX)} for the \textit{unit agent} to fill the current bottle \texttt{X} with the liquid type \texttt{LQ} within the filling level range \texttt{MN-MX}.
Once the bottle is filled, the norm \texttt{n2} is activated and produces an obligation for the \textit{unit agent} updating the deviation and learning factors based on the bottle filled.
If the polarity is negative and the image drops below a threshold, the norm becomes \texttt{unfulfilled} and
%
%\change[OB]{In this case, two sanctions may be applied.
%The sanction conditions allow the sanctions to be adapted according to the severity of the incorrect bottle filling.}{
one of two sanction facts is produced, i.e., \texttt{s1} and \texttt{s2}.
%conditional on the severity of the incorrect bottle filling.
%
If the number of consecutive violations (\texttt{C}) of the \textit{valve agent} (\texttt{V}) is less than a threshold \texttt{T}, then the sanction fact \texttt{s1} is produced ordering the \textit{valve agent} (\texttt{V}) to adjust its flow rate; otherwise, the sanction fact \texttt{s2} is produced ordering the \textit{valve agent} (\texttt{V}) to activate the self cleaning procedure in the \textit{valve} artifact.

Next the implementation of this sanctioning enforcement mechanism is shown in the context of the \textit{myJoghurt} MAS architecture.
\subsection{\textit{myJoghurt} MAS architecture}
%---------------------------------------------------------
%\note[EY]{New, please check. My idea is to divide in two subsections: one dealing with architecture and one with language - title to be redefined}
%\note[OB]{Do we use capability or component?}
%
\begin{figure}[ht]
      \centering
      \includegraphics[width=.75\textwidth]{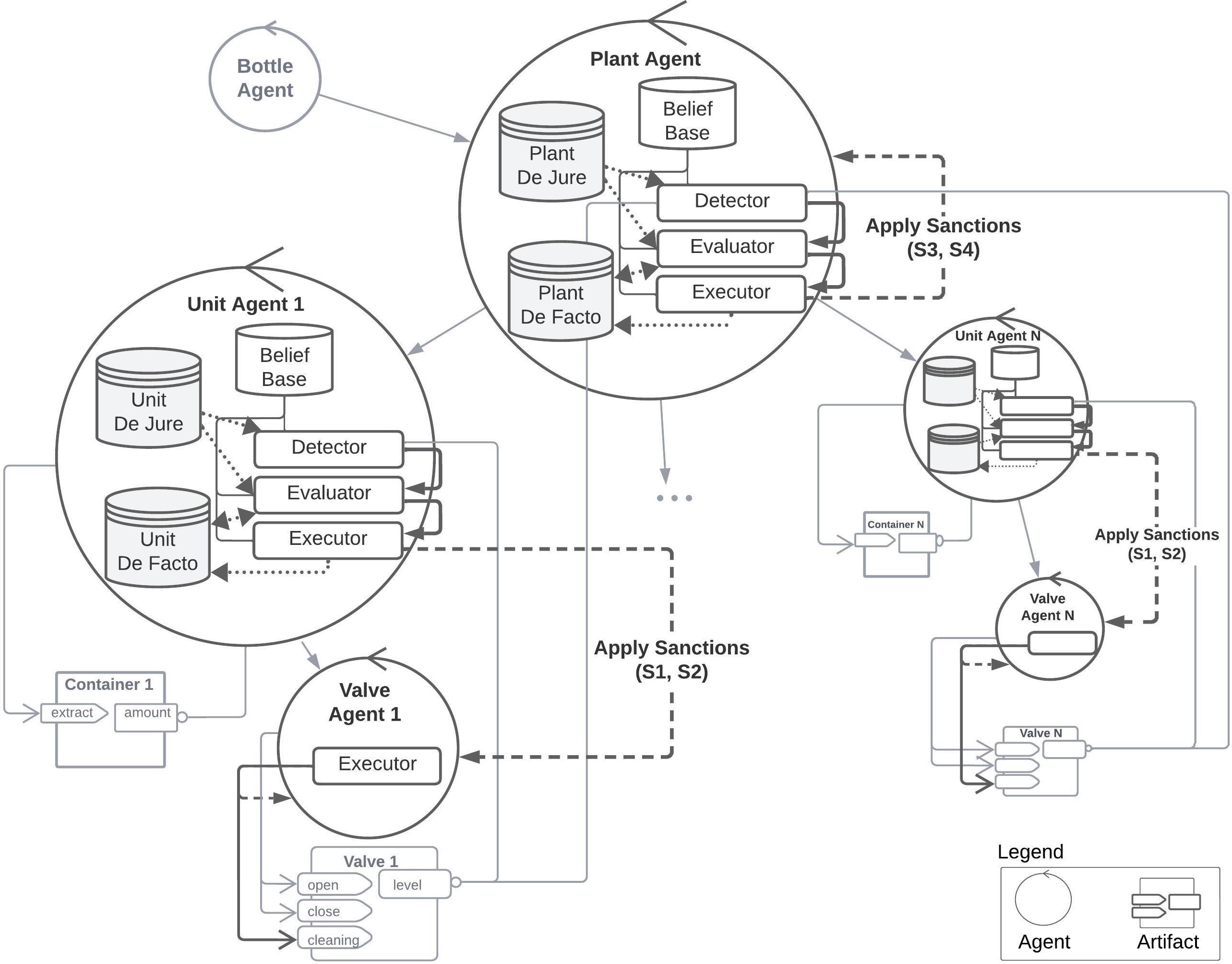}
      \caption{Overview of integration of enforcement mechanisms into the MAS Architecture.}
      \label{fig:architecture-sanctions}
\end{figure}
%
% \note[EY]{Or should I make one De Jure repository outside the agent and represent norms loaded inside the agent?}

The normative agent architecture % enforcement \annote[EY]{model}{mechanisms?} %\change[OB]{is incorporated into the multi-agent architecture (see Fig.~\ref{fig:architecture-sanctions}) following the proposal in Section~\ref{sec:architecture}}{
equipped with the NPL(s) engine proposed in Section~\ref{sec:architecture} is used to design agents in the \textit{myJoghurt} MAS architecture (Fig.~\ref{fig:architecture-sanctions}).
%to enhance the robustness of the production plant.
% \note[OB]{In what follows, I have doubts on the way this integration is done: is it distributed into agents? does it involve also artifacts? From the schema, it doesn't seem to be the case, but not sure. }
%
The \textit{plant agent} and \textit{unit agents} set the norms and sanctions in their De Jure repository and they are endowed with the enforcement.
%
%\annote[OB]{The De Jure is loaded internally into the agents, which store specific norms and the associated sanction rules for the unit and plant.}{delete this sentence since said above?}
%
Because the applied agent architecture is modular, we can adapt each agent to specific requirements of the \textit{myJoghurt} architecture.
%
% In this case study, the capabilities are \annote[EY]{implemented inside the agent.}{Capabilities can be realized in multiple ways, through artifacts or new components, but in our case, there are no external components, they are implemented inside the agent - please check}\note[OB]{I like the annotation that you did, maybe could be part of the text with the explanation of the rational behind the choice?}
%
% The unit agent implements the capabilities of Detector, Evaluator, and Executor, while the plant agent implements the capabilities of Evaluator and Executor.
%
%\remove[OB]{The technology that we used for realizing this architecture is JaCaMo~cite\{boissier2020multi\} with the integration of NPL(s) for the normative and sanction parts.}
%
So, the \textit{plant agent} and the \textit{unit agents} implement the capabilities of Detector, Evaluator, and Executor, while the \textit{valve agent} implements only the capability of Executor.

% DETECTOR
The Detector capability of the \textit{unit agent} and \textit{plant agent} perceives the filling level information of each completed bottle and detects any violation or compliance with the norms.
%
% The plant agent will also be informed of the result.
%
% The Detector capability of the bottle agent\remove[EY]{ or other specialized agent, e.g., one equipped with detection sensors,} monitors the filling level of the bottle and informs all \change[EY]{Evaluators}{Evaluator agents}.
%
% \note[OB]{what is Evaluator? a role? a component incorporated in an agent? if so say: all agents incorporating evaluators? but how does an agent know who has the capability of evaluating?}
%
% \change[EY]{Each bottle must be filled within its tolerance range.}{
%The norm defined in this use case is that each bottle must be filled within a specified tolerance range.
%} 
%\note[OB]{Is it a norm? if so tell it explicitly.}

% EVALUATOR
Once a violation is detected, both the \textit{plant agent} and \textit{unit agents} calculate and update the factors to determine whether and which sanctions to apply using their Evaluator capability.
The Evaluator considers the norms in the De Jure and the historical information in the De Facto to make these decisions.
Based on the \change[EY]{\textit{valve agent} image and the number of consecutive prior violations to \texttt{n1},}{sanctions identified in Table{~\ref{tab:sanctions}}}, the \textit{unit agent} may apply sanctions S1 and S2 to the \textit{valve agent} responsible for the violation. % \note[OB]{how do you know which valve agent is the respective one? kind of organisation?}
%\note[EY]{in the code actually refers only to a single group of agents, but maybe it should be handled with an organization}
%\note[OB]{I agree, but we don't have time to explore this. We can let it not said for now and will think about it later. This is a minor point.}
%
%These sanctions are classified as other-directed locus with guidance purpose~\cite{nardin2016sanctions}).
%
S2 and S3 are applied by the \textit{plant agent} to itself in order to indirectly update the \textit{unit agents}' image or even remove it from the considerations of future orders.
%(i.e., self-directed locus with incapacitation purpose).
%
S5 is not represented in the diagram since it is a sanction to be sent to an external actor in the system (e.g., a human operator) responsible for repairing the filling station.
%
% The Evaluator \change[EY]{model}{capability}\note[OB]{is it a model or a component?} model uses a set of factors to reason and decide whether and which sanctions to apply, in particular 

% EXECUTOR
The Executor capabilities of \textit{plant agent}, \textit{unit agents}, and \textit{valve agent} %(\change[EY]{enacted}{implemented}\note[OB]{enacted: this term may be confusing, since we use it usually to talk about roles, norms. use "implemented" instead?} by the unit and plant agents) 
handle the actual enforcement of the sanctions.
%
% the executor is obedient -> they don't take any decision
We assume Executors are obedient and they enforce the sanctions chosen by the Evaluator. However, there may be cases in which the Executor may decide whether or not to enforce the sanctions.
The actual sanction enforcement is specified in the agent code as a plan.
%
% reaction to the creation of the sanction
The following listing shows the \textit{unit agent} internal plan for reacting to the sanction fact of sanction S1 linked to adjusting the \textit{valve agent} flow rate estimation.
\begin{lstlisting}[
    language={},
    label={lst:sanction}
]
+sanction(V, adjust_flow_rate(X))
    <-  //...
        .send(V, signal, update_estimation(M)).
\end{lstlisting}
%
%\add[EY]{It is important to underline that, because of the flexibility in the agent-centric perspective, agents may exhibit diverse reactions to the same sanction.}

For the sake of simplicity in this paper, we have not considered the Controller and Legislator capabilities.
%==================================================================================
\section{Conclusion and Future Work}\label{sec:conclusion}
%==================================================================================
%\note[EY]{future work: combining institutional approach and agent approach, possibility of having conflicting norms and managing them, support for transparency and explainability, complete the case study and test it}
%
%
In this paper, we have presented an agent-centric perspective on norm enforcement and sanctions by proposing a twofold contribution. 
We have extended the normative programming language NPL to NPL(s) with an explicit representation of sanctions as first-class abstraction in the definition of a normative program. 
Thanks to it, normative programs gain flexibility and expressivity enabling the association of sanctions with different norms on one hand, and, on the second hand, open the possibility for agents to locally decide on sanctions and how to enforce them.

Based on this enriched normative programming language, we proposed a normative agent architecture able to reason explicitly on norms and sanctions. 
The use of these new features has been illustrated in a multi-agent based control production system in the context of the industrial case study of the laboratory plant \emph{myJoghurt}.
We have shown how this normative agent-centric architecture opens a comprehensive, modular, and flexible perspective on norm enforcement and sanctions in a multi-agent system.
Future work is to experiment the case study in the real setting of the \emph{myJoghurt} plant to assess the enforcement approach and to evaluate the agent's performance. 
From a theoretical point of view, our proposal opens the path to explore transparency and explainability of the normative functioning of multi-agent systems taking profit of this agent-centric reasoning on norms and sanctions extending the approach proposed in~\cite{yan2023explainability}. 
\add[EY]{Future extension is to consider the distinctions between different types of obligations (i.e., an obligation as an action to be performed and/or to achieve a certain state) and how agents adapt their sanctioning strategy effectively, especially in complex social scenarios.}
Another extension concerns the connection of our sanctioning mechanisms with the approach developed in Situated Artificial Institution by connecting the sanctioning process directly to the environment~\cite{brito19}.
\section*{Acknowledgment}
Partially funded by ANR-FAPESP NAIMAN project (ANR-22-CE23-0018-01), GFA through RAMP-UP II project, and \add[JH]{CAPES, The Brazilian Agency for Higher Education, under the project PrInt CAPES-UFSC ``Automation 4.0''}.

\let\OLDthebibliography\thebibliography

\renewcommand\thebibliography[1]{
  \OLDthebibliography{#1}
  \fontsize{8}{9}\selectfont
  \setlength{\parskip}{0pt}
  \setlength{\itemsep}{0pt plus 0.3ex}
}

\bibliographystyle{splncs04}
\bibliography{bibliography,jomi}
\end{document}